\documentclass{iauc}
\usepackage{graphicx}

\title[Virgo dEs in the NIR] %% give here short title %% 
{NIR observations of dEs
in the Virgo cluster: a structural continuity with giant Ellipticals}
\author[S.~Zibetti]   %% give here short author list %%
{Stefano Zibetti$^{1}$ \and the GOLDMiNe Research Team$^{2}$
}

\affiliation{$^1$Max-Planck-Institut f\"ur Extraterrestrische Physik \break
Gie\ss enbachstra\ss e, D-85748, Garching bei M\"unchen, Germany $-$ 
email: szibetti@mpe.mpg.de \break
$^2$\texttt{http://goldmine.mib.infn.it/}}

\pubyear{2005}
\volume{198} %% insert here IAU Symposium No.
\pagerange{0--1}
\date{?? and in revised form ??}
\jname{Proceedings Title IAU Colloquium }
\editors{B. Binggeli, and H. Jerjen, eds.}
\begin{document}

\maketitle

\begin{abstract} The structural properties of a sample of 50 dEs in the Virgo cluster are here
derived from Near InfraRed (NIR, H-band 1.65 $\mu$m) surface photometry and analysed.
One-dimensional surface
brightness (SB) profiles are extracted using elliptical isophote fitting. They are characterised by
means of structural parameters, namely the half light radius $R_e$, the average surface brightness
within $R_e$ ($\mu_e$), and a concentration index ($c_{31}$). We show that typical dEs have
close-to-exponential NIR SB distributions.\\
The relations between dEs and giant ellipticals (Es) are investigated by comparing the NIR
structural parameters of 273 Es in nearby clusters. Further analysis is conducted using the
optical-NIR colour $B-H$ and by studying the relationships between structural and dynamical
parameters (fundamental plane) for the two classes of galaxies. The transition between the two
regimes is smooth and no dichotomy is seen.

%% add here a maximum of 10 keywords, to be taken form the file <Keywords.txt>.

\end{abstract}

\firstsection % if your document starts with a section,
\section{Introduction and sample} 
Dwarf elliptical galaxies are {\bf not} just a rescaled version of their giant counterparts.
Optical observations have shown that, as opposed to Es, dEs have exponential rather than de
Vaucouleurs SB profiles, and do not follow the same scaling relations linking luminosity, the
effective radius $R_e$ and the mean effective surface brightness $\mu_e$.\\ In this work we
re-assess the structural properties of dEs using NIR surface photometry, which is a good tracer of
the bulk of the stellar mass, while being much less affected than optical bands by dust attenuation
and metallicity effects. This project is part of a large NIR survey of galaxies in nearby clusters
\cite[(Gavazzi et al. 2000, 2003)] {peppo_franz,goldmine}. This has allowed to compare the derived
properties with those of E galaxies in the full range of luminosities.

The 50 dEs of this work have been selected from the VCC \cite[(Binggeli, Sandage \& Tammann,
1985)]{VCC}, limited to photographic magnitude $m_{\mathrm p}<16$, and observed at the 3.6m
telescopes ESO/NTT and TNG \cite[(Gavazzi et al. 2001)]{peppo_ziby}. Our sample is representative
of $\sim$30\% of the dEs more luminous than $10^{8.5}$L$_\odot$ (H). 48 out of these 50 have B-band
photometry available, and 6 have also velocity dispersion measurements. 273 Es (including 35 in
Virgo and the complete sample of 217  Es in Coma and Abell 1367) have been used to investigate the
relations between dwarves and giants.
\section{Results}
The SB profile of each galaxy is extracted by fitting elliptical isophotes with standard IRAF
procedures. The decomposition of the 1-D profile with simple analytical functions  (exponential, de
Vaucouleurs and their combinations) is used to extrapolate an asymptotic magnitude (extrapolations
are of the order of 0.1--0.2 mag). \emph{From the measured SB profile} we extract the effective
radius $R_e$ enclosing half of the asymptotic flux, and the average SB $\mu_e$ within $R_e$. The
distribution of dEs and Es in this plane (Fig. 1a) does not display any significant gap between the
two populations. The ``Kormendy'' relation for Es (thick dashed line) represents the upper limit
for the SB at given $R_e$, with dEs having lower SB than Es. The multi-dimensional photometric
parameter space $c_{31}-(B-H)-L_H$\footnote{The concentration index $c_{31}$ is the ratio of the
radii enclosing 75\% and 25\%  of the asymptotic flux.} \cite[(Fig 1b, see also Scodeggio et al.
2002)]{cubo_marco} shows very clearly the smooth transition from blue, low-concentration (nearly
exponential) dEs to red,  concentrated (roughly de Vaucouleurs) Es.
\begin{figure}
\begin{center}
\includegraphics[width=0.49\textwidth]{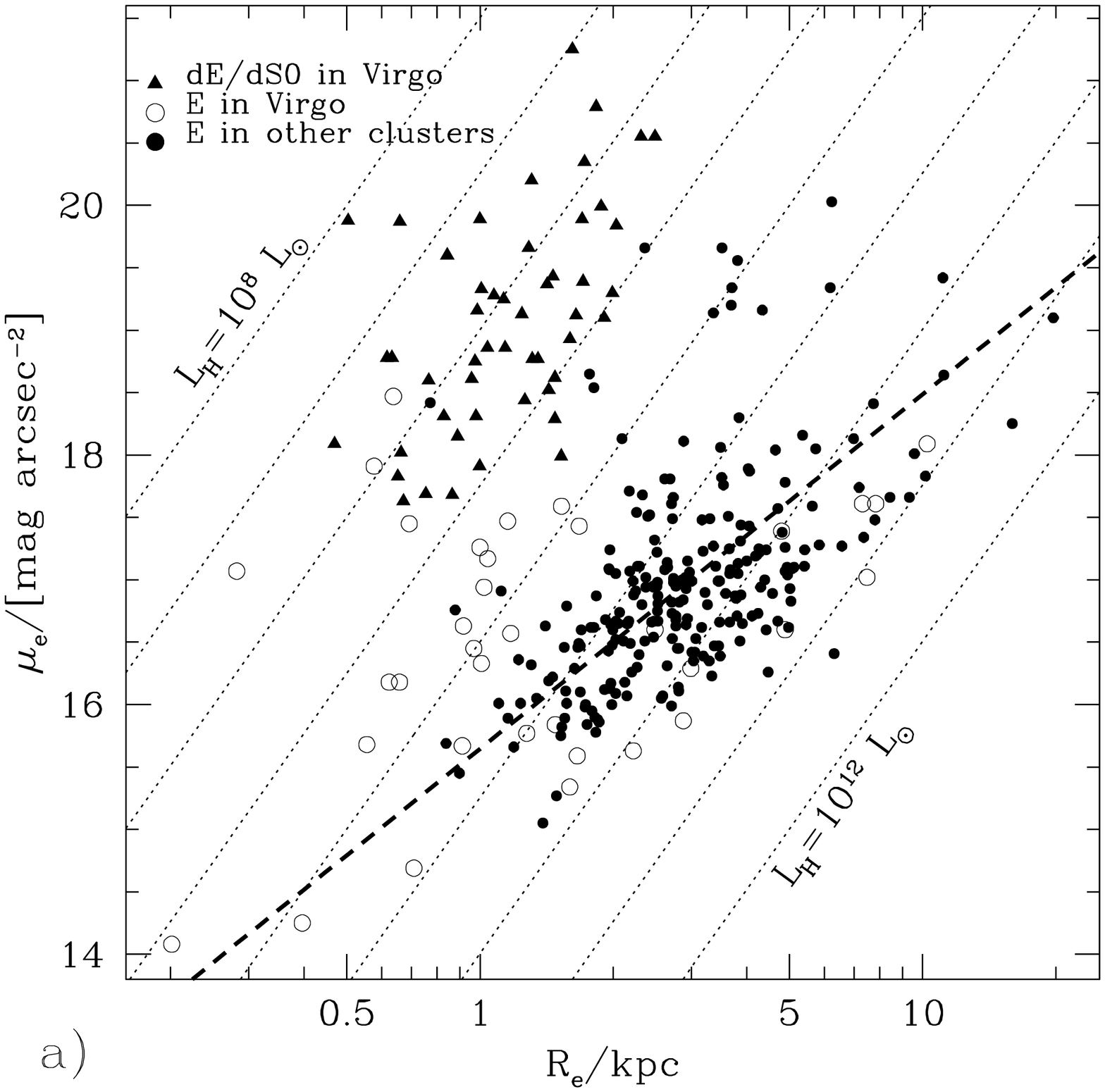}
\includegraphics[width=0.49\textwidth]{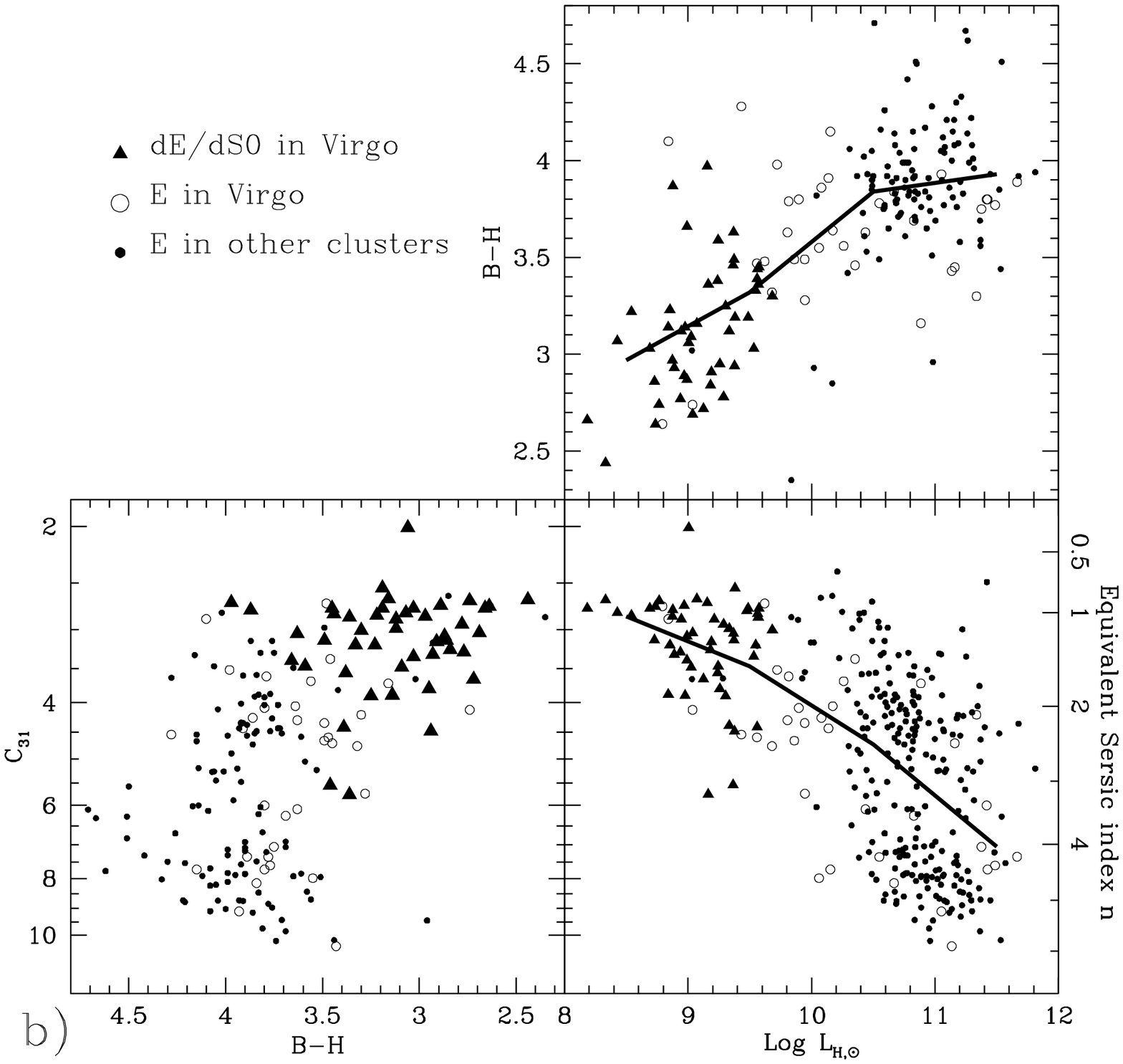}
\end{center}
\caption{{\bf a)} The $H$-band $\mu_e-R_e$ plane. 
The thick dashed line is the ``Kormendy'' relation for the Es. Dotted lines are isoluminosity lines
assuming a fixed ellipticity=0.35. {\bf b)} The ``photometric cube'', alias $c_{31}-(B-H)-L_H$
space. Trends are indicated by the thick lines, representing the median of $c_{31}$  and $B-H$ in 4
bins of $L_H$. The right-side scale of the $c_{31}-L_H$ panel reports the value of the S\'ersic
index $n$ that reproduces the same $c_{31}$.}
\end{figure}

We have analysed the Fundamental Plane relations for the galaxies with a velocity dispersion 
measure \cite{ziby02}. Using NIR structural parameters, dEs lie on the same 
relations as Es.
Systematic variations of $M/L$ in the NIR have been investigated by means of the 
$\kappa$-space formalism \cite[(Bender, Burstein \& Faber, 1992)]{k3}: although a trend with mass is
clearly seen at high masses, dEs span the same range in $M/L$ as low-mass Es.
Adopting a more sophisticated formalism,
in which profile deprojection is used to take non-homology and spectroscopic aperture effects
into account \cite[(Zibetti et al. 2002)]{ziby02}, we show that the data are consistent with 
constant $M/L$ throughout the sequence of Es and dEs.

\end{document}